\input harvmac
\def\ZZ{\hbox{Z\kern-.4emZ}}
\def\RR{\hbox{R\kern-.6emR}}

\lref\BredbergPV{
  I.~Bredberg, T.~Hartman, W.~Song and A.~Strominger,
  ``Black Hole Superradiance From Kerr/CFT,''
  arXiv:0907.3477 [hep-th].
}

\lref\BrownNW{
  J.~D.~Brown and M.~Henneaux,
 ``Central Charges in the Canonical Realization of Asymptotic Symmetries: An Example from Three-Dimensional Gravity,''
Commun.\ Math.\ Phys.\ \ {\bf 104}, 207  (1986)..
}

\lref\LarsenXM{
  F.~Larsen,
  ``The Attractor Mechanism in Five Dimensions,''
Lect.\ Notes Phys.\  {\bf 755}, 249-281 (2008)
[hep-th/0608191].
}

\lref\CveticZQ{
  M.~Cveti\v c and C.~M.~Hull,
  ``Black holes and U-duality,''
  Nucl.\ Phys.\  B {\bf 480}, 296 (1996)
  [arXiv:hep-th/9606193].
}

\lref\BrownNW{
  J.~D.~Brown and M.~Henneaux,
  ``Central Charges in the Canonical Realization of Asymptotic Symmetries: An Example from Three-Dimensional Gravity,''
Commun.\ Math.\ Phys.\  {\bf 104}, 207-226 (1986).
}

\lref\MisnerKX{
  C.~W.~Misner,
  ``Interpretation of gravitational-wave observations,''
  Phys.\ Rev.\ Lett.\  {\bf 28}, 994 (1972).
}
\lref\TeukolskyHA{
  S.~A.~Teukolsky,
  ``Perturbations of a rotating black hole. 1. Fundamental equations for
  gravitational electromagnetic and neutrino field perturbations,''
  Astrophys.\ J.\  {\bf 185}, 635 (1973).
}

\lref\PressZZ{
  W.~H.~Press and S.~A.~Teukolsky,
  ``Perturbations of a Rotating Black Hole. II. Dynamical Stability of the Kerr
  Metric,''
  Astrophys.\ J.\  {\bf 185}, 649 (1973).
}

\lref\FrolovJH{
  V.~P.~Frolov and K.~S.~Thorne,
  Phys.\ Rev.\  D {\bf 39}, 2125 (1989).
}
\lref\CveticUW{
  M.~Cveti\v c and F.~Larsen,
  ``General rotating black holes in string theory: Greybody factors and  event
  horizons,''
  Phys.\ Rev.\  D {\bf 56}, 4994 (1997)
  [arXiv:hep-th/9705192].
}

\lref\CveticXV{
  M.~Cveti\v c and F.~Larsen,
  ``Greybody factors for rotating black holes in four dimensions,''
  Nucl.\ Phys.\  B {\bf 506}, 107 (1997)
  [arXiv:hep-th/9706071].
}

\lref\CveticVP{
  M.~Cveti\v c and F.~Larsen,
  ``Black hole horizons and the thermodynamics of strings,''
  Nucl.\ Phys.\ Proc.\ Suppl.\  {\bf 62}, 443 (1998)
  [Nucl.\ Phys.\ Proc.\ Suppl.\  {\bf 68}, 55 (1998)]
  [arXiv:hep-th/9708090].
}
\lref\CveticAP{
  M.~Cveti\v c and F.~Larsen,
  ``Greybody factors for black holes in four dimensions: Particles with
  spin,''
  Phys.\ Rev.\  D {\bf 57}, 6297 (1998)
  [arXiv:hep-th/9712118].
}

\lref\LarsenGE{
  F.~Larsen,
  ``A string model of black hole microstates,''
  Phys.\ Rev.\  D {\bf 56}, 1005 (1997)
  [arXiv:hep-th/9702153].
}

\lref\DiasNJ{
  O.~J.~C.~Dias, R.~Emparan and A.~Maccarrone,
  ``Microscopic Theory of Black Hole Superradiance,''
  Phys.\ Rev.\  D {\bf 77}, 064018 (2008)
  [arXiv:0712.0791 [hep-th]].
}

\lref\BredbergPV{
  I.~Bredberg, T.~Hartman, W.~Song and A.~Strominger,
  ``Black Hole Superradiance From Kerr/CFT,''
  arXiv:0907.3477 [hep-th].
}

\lref\StromingerSH{
  A.~Strominger and C.~Vafa,
  ``Microscopic Origin of the Bekenstein-Hawking Entropy,''
  Phys.\ Lett.\  B {\bf 379}, 99 (1996)
  [arXiv:hep-th/9601029].
}

\lref\CastroMS{
  A.~Castro, D.~Grumiller, F.~Larsen and R.~McNees,
  ``Holographic Description of AdS$_2$ Black Holes,''
  JHEP {\bf 0811}, 052 (2008)
  [arXiv:0809.4264 [hep-th]];
  A.~Castro and F.~Larsen,
  ``Near Extremal Kerr Entropy from AdS_2 Quantum Gravity,''
  arXiv:0908.1121 [hep-th].
}
\lref\HartmanDQ{
  T.~Hartman and A.~Strominger,
  ``Central Charge for AdS$_2$ Quantum Gravity,''
  JHEP {\bf 0904}, 026 (2009)
  [arXiv:0803.3621 [hep-th]].
}

\lref\GuicaMU{
  M.~Guica, T.~Hartman, W.~Song and A.~Strominger,
  ``The Kerr/CFT Correspondence,''
  arXiv:0809.4266 [hep-th].
}

\lref\SenQY{
  A.~Sen,
  ``Black Hole Entropy Function, Attractors and Precision Counting of
  Microstates,''
  Gen.\ Rel.\ Grav.\  {\bf 40}, 2249 (2008)
  [arXiv:0708.1270 [hep-th]].
}

\lref\BardeenPX{
  J.~M.~Bardeen and G.~T.~Horowitz,
  ``The extreme Kerr throat geometry: A vacuum analog of AdS(2) x S(2),''
  Phys.\ Rev.\  D {\bf 60}, 104030 (1999)
  [arXiv:hep-th/9905099].
}

\lref\BalasubramanianKQ{
  V.~Balasubramanian, A.~Naqvi and J.~Simon,
  ``A multi-boundary AdS orbifold and DLCQ holography: A universal  holographic
  description of extremal black hole horizons,''
  JHEP {\bf 0408}, 023 (2004)
  [arXiv:hep-th/0311237].
}
\lref\BalasubramanianBG{
  V.~Balasubramanian, J.~de Boer, M.~M.~Sheikh-Jabbari and J.~Simon,
  ``What is a chiral 2d CFT? And what does it have to do with extremal black
  holes?,''
  arXiv:0906.3272 [hep-th].
}

\lref\BarnichBF{
  G.~Barnich and G.~Compere,
  ``Surface charge algebra in gauge theories and thermodynamic integrability,''
  J.\ Math.\ Phys.\  {\bf 49}, 042901 (2008)
  [arXiv:0708.2378 [gr-qc]].
}
\lref\CompereAZ{
  G.~Compere,
  ``Symmetries and conservation laws in Lagrangian gauge theories with
  applications to the mechanics of black holes and to gravity in three
  dimensions,''
  arXiv:0708.3153 [hep-th].
}
\lref\BarnichKQ{
  G.~Barnich and G.~Compere,
  ``Conserved charges and thermodynamics of the spinning Goedel black hole,''
  Phys.\ Rev.\ Lett.\  {\bf 95}, 031302 (2005)
  [arXiv:hep-th/0501102].
}
\lref\BanadosDA{
  M.~Banados, G.~Barnich, G.~Compere and A.~Gomberoff,
  ``Three dimensional origin of Goedel spacetimes and black holes,''
  Phys.\ Rev.\  D {\bf 73}, 044006 (2006)
  [arXiv:hep-th/0512105].
}

\lref\EmparanEN{
  R.~Emparan and A.~Maccarrone,
  ``Statistical Description of Rotating Kaluza-Klein Black Holes,''
  Phys.\ Rev.\  D {\bf 75}, 084006 (2007)
  [arXiv:hep-th/0701150].
}

\lref\DiasNJ{
  O.~J.~C.~Dias, R.~Emparan and A.~Maccarrone,
  ``Microscopic Theory of Black Hole Superradiance,''
  Phys.\ Rev.\  D {\bf 77}, 064018 (2008)
  [arXiv:0712.0791 [hep-th]].
}

\lref\BardeenPX{
  J.~M.~Bardeen and G.~T.~Horowitz,
  ``The extreme Kerr throat geometry: A vacuum analog of AdS(2) x S(2),''
  Phys.\ Rev.\  D {\bf 60}, 104030 (1999)
  [arXiv:hep-th/9905099].
}

\lref\BalasubramanianRE{
  V.~Balasubramanian and P.~Kraus,
  ``A stress tensor for anti-de Sitter gravity,''
  Commun.\ Math.\ Phys.\  {\bf 208}, 413 (1999)
  [arXiv:hep-th/9902121].
}

\lref\SkenderisWP{
  K.~Skenderis,
  ``Lecture notes on holographic renormalization,''
  Class.\ Quant.\ Grav.\  {\bf 19}, 5849 (2002)
  [arXiv:hep-th/0209067].
}

\lref\AmselEV{
  A.~J.~Amsel, G.~T.~Horowitz, D.~Marolf and M.~M.~Roberts,
  ``No Dynamics in the Extremal Kerr Throat,''
  arXiv:0906.2376 [hep-th].
}
\lref\DiasEX{
  O.~J.~C.~Dias, H.~S.~Reall and J.~E.~Santos,
  ``Kerr-CFT and gravitational perturbations,''
  arXiv:0906.2380 [hep-th].
}

\lref\MaldacenaBW{
  J.~M.~Maldacena and A.~Strominger,
  ``AdS(3) black holes and a stringy exclusion principle,''
  JHEP {\bf 9812}, 005 (1998)
  [arXiv:hep-th/9804085].
}

\lref\DijkgraafFQ{
  R.~Dijkgraaf, J.~M.~Maldacena, G.~W.~Moore and E.~P.~Verlinde,
  ``A black hole farey tail,''
  arXiv:hep-th/0005003.
}

\lref\KrausVZ{
  P.~Kraus and F.~Larsen,
  ``Microscopic Black Hole Entropy in Theories with Higher Derivatives,''
  JHEP {\bf 0509}, 034 (2005)
  [arXiv:hep-th/0506176].
}

\lref\LarsenBU{
  F.~Larsen,
  ``Anti-de Sitter spaces and nonextreme black holes,''
  arXiv:hep-th/9806071.
}

\lref\CveticXV{
  M.~Cveti\v c and F.~Larsen,
  ``Greybody factors for rotating black holes in four dimensions,''
  Nucl.\ Phys.\  B {\bf 506}, 107 (1997)
  [arXiv:hep-th/9706071].
}

\lref\MaldacenaDS{
  J.~M.~Maldacena and L.~Susskind,
  ``D-branes and Fat Black Holes,''
  Nucl.\ Phys.\  B {\bf 475}, 679 (1996)
  [arXiv:hep-th/9604042].
}
\lref\KunduriVF{
  H.~K.~Kunduri, J.~Lucietti and H.~S.~Reall,
  ``Near-horizon symmetries of extremal black holes,''
  Class.\ Quant.\ Grav.\  {\bf 24}, 4169 (2007)
  [arXiv:0705.4214 [hep-th]].
}

\lref\StromingerYG{
  A.~Strominger,
  ``AdS(2) quantum gravity and string theory,''
  JHEP {\bf 9901}, 007 (1999)
  [arXiv:hep-th/9809027].
}
\lref\ChoFZ{
  J.~H.~Cho, T.~Lee and G.~W.~Semenoff,
  ``Two dimensional anti-de Sitter space and discrete light cone
  quantization,''
  Phys.\ Lett.\  B {\bf 468}, 52 (1999)
  [arXiv:hep-th/9906078].
}

\lref\GimonUR{
  E.~G.~Gimon and P.~Ho\v rava,
  ``Astrophysical Violations of the Kerr Bound as a Possible Signature of
  String Theory,''
  Phys.\ Lett.\  B {\bf 672}, 299 (2009)
  [arXiv:0706.2873 [hep-th]].
}
\lref\AzeyanagiBJ{
  T.~Azeyanagi, T.~Nishioka and T.~Takayanagi,
  ``Near Extremal Black Hole Entropy as Entanglement Entropy via AdS2/CFT1,''
  Phys.\ Rev.\  D {\bf 77}, 064005 (2008)
  [arXiv:0710.2956 [hep-th]].
}

\lref\GuptaKI{
  R.~K.~Gupta and A.~Sen,
  ``Ads(3)/CFT(2) to Ads(2)/CFT(1),''
  JHEP {\bf 0904}, 034 (2009)
  [arXiv:0806.0053 [hep-th]].
}

\lref\StromingerEQ{
  A.~Strominger,
  ``Black hole entropy from near horizon microstates,''
JHEP {\bf 9802}, 009 (1998)
[hep-th/9712251].
}

\lref\MaldacenaUZ{
  J.~M.~Maldacena, J.~Michelson and A.~Strominger,
  ``Anti-de Sitter fragmentation,''
  JHEP {\bf 9902}, 011 (1999)
  [arXiv:hep-th/9812073].
}

\lref\MaldacenaIH{
  J.~M.~Maldacena and A.~Strominger,
  ``Universal low-energy dynamics for rotating black holes,''
  Phys.\ Rev.\  D {\bf 56}, 4975 (1997)
  [arXiv:hep-th/9702015].
}
\lref\MathurET{
  S.~D.~Mathur,
  ``Absorption of angular momentum by black holes and D-branes,''
  Nucl.\ Phys.\  B {\bf 514}, 204 (1998)
  [arXiv:hep-th/9704156].
}
\lref\GubserQR{
  S.~S.~Gubser,
  ``Can the effective string see higher partial waves?,''
  Phys.\ Rev.\  D {\bf 56}, 4984 (1997)
  [arXiv:hep-th/9704195].
}

\lref\LarsenXM{
  F.~Larsen,
  ``The attractor mechanism in five dimensions,''
  Lect.\ Notes Phys.\  {\bf 755}, 249 (2008)
  [arXiv:hep-th/0608191].
}
\lref\CveticXZ{
  M.~Cveti\v c and D.~Youm,
  ``General Rotating Five Dimensional Black Holes of Toroidally Compactified
  Heterotic String,''
  Nucl.\ Phys.\  B {\bf 476}, 118 (1996)
  [arXiv:hep-th/9603100].
}

\lref\CveticKV{
  M.~Cveti\v c and D.~Youm,
  ``Entropy of Non-Extreme Charged Rotating Black Holes in String Theory,''
  Phys.\ Rev.\  D {\bf 54}, 2612 (1996)
  [arXiv:hep-th/9603147].
}

\lref\HartmanPB{
  T.~Hartman, K.~Murata, T.~Nishioka and A.~Strominger,
  ``CFT Duals for Extreme Black Holes,''
  JHEP {\bf 0904}, 019 (2009)
  [arXiv:0811.4393 [hep-th]].
}

\lref\LarsenGE{
  F.~Larsen,
  ``A String model of black hole microstates,''
Phys.\ Rev.\  {\bf D56}, 1005-1008 (1997)
[hep-th/9702153].
}

\lref\MaldacenaIX{
  J.~M.~Maldacena and A.~Strominger,
  ``Black hole greybody factors and D-brane spectroscopy,''
  Phys.\ Rev.\  D {\bf 55}, 861 (1997)
  [arXiv:hep-th/9609026].
}

\lref\MaldacenaIH{
  J.~M.~Maldacena and A.~Strominger,
  ``Universal low-energy dynamics for rotating black holes,''
  Phys.\ Rev.\  D {\bf 56}, 4975 (1997)
  [arXiv:hep-th/9702015].
}
\lref\CallanTV{
  C.~G.~.~Callan, S.~S.~Gubser, I.~R.~Klebanov and A.~A.~Tseytlin,
  ``Absorption of fixed scalars and the D-brane approach to black holes,''
  Nucl.\ Phys.\  B {\bf 489}, 65 (1997)
  [arXiv:hep-th/9610172].
}
\lref\GubserZP{
  S.~S.~Gubser and I.~R.~Klebanov,
  ``Four-dimensional greybody factors and the effective string,''
  Phys.\ Rev.\ Lett.\  {\bf 77}, 4491 (1996)
  [arXiv:hep-th/9609076].
}
\lref\ChowDP{
  D.~D.~K.~Chow, M.~Cveti\v c, H.~L\"u and C.~N.~Pope,
  ``Extremal Black Hole/CFT Correspondence in (Gauged) Supergravities,''
  arXiv:0812.2918 [hep-th].
}

\lref\LopesCardosoKY{
  G.~Lopes Cardoso, A.~Ceresole, G.~Dall'Agata, J.~M.~Oberreuter and J.~Perz,
 ``First-order flow equations for extremal black holes in very special
 geometry,''
  JHEP {\bf 0710}, 063 (2007)
  [arXiv:0706.3373 [hep-th]].
}

\lref\BalasubramanianEE{
  V.~Balasubramanian and F.~Larsen,
  ``Near horizon geometry and black holes in four-dimensions,''
Nucl.\ Phys.\  {\bf B528}, 229-237 (1998)
[hep-th/9802198].
}

\lref\HottaWZ{
  K.~Hotta and T.~Kubota,
  ``Exact Solutions and the Attractor Mechanism in Non-BPS Black Holes,''
  Prog.\ Theor.\ Phys.\  {\bf 118}, 969 (2007)
  [arXiv:0707.4554 [hep-th]].
}
\lref\GimonGK{
  E.~G.~Gimon, F.~Larsen and J.~Simon,
  ``Black Holes in Supergravity: the non-BPS Branch,''
  JHEP {\bf 0801}, 040 (2008)
  [arXiv:0710.4967 [hep-th]].
  ``Constituent Model of Extremal non-BPS Black Holes,''
  JHEP {\bf 0907}, 052 (2009)
  [arXiv:0903.0719 [hep-th]].
  }
\lref\BenaEV{
  I.~Bena, G.~Dall'Agata, S.~Giusto, C.~Ruef and N.~P.~Warner,
  ``Non-BPS Black Rings and Black Holes in Taub-NUT,''
  JHEP {\bf 0906}, 015 (2009)
  [arXiv:0902.4526 [hep-th]].
}
\lref\deBoerIP{
  J.~de Boer,
  ``Six-dimensional supergravity on S**3 x AdS(3) and 2d conformal field
  theory,''
  Nucl.\ Phys.\  B {\bf 548}, 139 (1999)
  [arXiv:hep-th/9806104].
}

\lref\KrausVZ{
  P.~Kraus and  F.~Larsen,
  ``Microscopic black hole entropy in theories with higher derivatives,''
JHEP {\bf 0509}, 034 (2005).
[hep-th/0506176].
}

\lref\KastorGT{
  D.~Kastor and J.~H.~Traschen,
  ``A very effective string model?,''
  Phys.\ Rev.\  D {\bf 57}, 4862 (1998)
  [arXiv:hep-th/9707157].
}

\lref\DasWN{
  S.~R.~Das and S.~D.~Mathur,
  ``Comparing decay rates for black holes and D-branes,''
  Nucl.\ Phys.\  B {\bf 478}, 561 (1996)
  [arXiv:hep-th/9606185].
}

\lref\PeetES{
  A.~W.~Peet,
  ``The Bekenstein formula and string theory (N-brane theory),''
  Class.\ Quant.\ Grav.\  {\bf 15}, 3291 (1998)
  [arXiv:hep-th/9712253].
}
\lref\DavidWN{
  J.~R.~David, G.~Mandal and S.~R.~Wadia,
  ``Microscopic formulation of black holes in string theory,''
  Phys.\ Rept.\  {\bf 369}, 549 (2002)
  [arXiv:hep-th/0203048].
}
\lref\SenQY{
  A.~Sen,
  ``Black Hole Entropy Function, Attractors and Precision Counting of
  Microstates,''
  Gen.\ Rel.\ Grav.\  {\bf 40}, 2249 (2008)
  [arXiv:0708.1270 [hep-th]].
}

\lref\PiolinePF{
  B.~Pioline and J.~Troost,
  ``Schwinger pair production in AdS(2),''
  JHEP {\bf 0503}, 043 (2005)
  [arXiv:hep-th/0501169].
}
\lref\KimXV{
  S.~P.~Kim and D.~N.~Page,
  ``Schwinger Pair Production in $dS_2$ and $AdS_2$,''
  Phys.\ Rev.\  D {\bf 78}, 103517 (2008)
  [arXiv:0803.2555 [hep-th]].
}
\lref\AzeyanagiDK{
  T.~Azeyanagi, N.~Ogawa and S.~Terashima,
  ``The Kerr/CFT Correspondence and String Theory,''
  Phys.\ Rev.\  D {\bf 79}, 106009 (2009)
  [arXiv:0812.4883 [hep-th]].
}

\lref\BarnichJY{
  G.~Barnich and F.~Brandt,
  ``Covariant theory of asymptotic symmetries, conservation laws and  central
  charges,''
  Nucl.\ Phys.\  B {\bf 633}, 3 (2002)
  [arXiv:hep-th/0111246].
}
\lref\CompereIN{
  G.~Compere and S.~Detournay,
  ``Centrally extended symmetry algebra of asymptotically Goedel spacetimes,''
  JHEP {\bf 0703}, 098 (2007)
  [arXiv:hep-th/0701039].
}
\lref\CveticJA{
  M.~Cveti\v c and F.~Larsen,
  ``Statistical entropy of four-dimensional rotating black holes from
  near-horizon geometry,''
  Phys.\ Rev.\ Lett.\  {\bf 82}, 484 (1999)
  [arXiv:hep-th/9805146].
}
\lref\CveticXH{
  M.~Cveti\v c and F.~Larsen,
  ``Near horizon geometry of rotating black holes in five dimensions,''
  Nucl.\ Phys.\  B {\bf 531}, 239 (1998)
  [arXiv:hep-th/9805097].
}

\lref\ChongZX{
  Z.~W.~Chong, M.~Cveti\v c, H.~L\"u and C.~N.~Pope,
  ``Non-extremal rotating black holes in five-dimensional gauged
  supergravity,''
  Phys.\ Lett.\  B {\bf 644}, 192 (2007)
  [arXiv:hep-th/0606213].
}

\lref\CveticJN{
  M.~Cveti\v c and F.~Larsen,
  ``Greybody Factors and Charges in Kerr/CFT,''
  JHEP {\bf 0909}, 088 (2009)
  [arXiv:0908.1136 [hep-th]].
}

\lref\CastroFD{
  A.~Castro, A.~Maloney and A.~Strominger,
  ``Hidden Conformal Symmetry of the Kerr Black Hole,''
  Phys.\ Rev.\  D {\bf 82}, 024008 (2010)
  [arXiv:1004.0996 [hep-th]].
}

\lref\MaldacenaIX{
  J.~M.~Maldacena and A.~Strominger,
  ``Black hole grey body factors and d-brane spectroscopy,''
Phys.\ Rev.\  {\bf D55}, 861-870 (1997)
[hep-th/9609026].
}

\lref\BredbergHP{
  I.~Bredberg, C.~Keeler, V.~Lysov and A.~Strominger,
  ``Cargese Lectures on the Kerr/CFT Correspondence,''
[arXiv:1103.2355 [hep-th]].
}

\lref\CveticMN{
  M.~Cveti\v c, G.~W.~Gibbons and C.~N.~Pope,
  ``Universal Area Product Formulae for Rotating and Charged Black Holes in Four and Higher Dimensions,''
  Phys.\ Rev.\ Lett.\  {\bf 106}, 121301 (2011)
[arXiv:1011.0008 [hep-th]].
}

\lref\MaldacenaDS{
  J.~M.~Maldacena and L.~Susskind,
  ``D-branes and fat black holes,''
  Nucl.\ Phys.\  B {\bf 475}, 679 (1996)
  [arXiv:hep-th/9604042].
}

\lref\KrausWN{
  P.~Kraus,
  ``Lectures on black holes and the AdS(3) / CFT(2) correspondence,''
Lect.\ Notes Phys.\  {\bf 755}, 193-247 (2008).
[hep-th/0609074].
}

\lref\KachruYH{
  S.~Kachru, X.~Liu and M.~Mulligan,
  ``Gravity Duals of Lifshitz-like Fixed Points,''
Phys.\ Rev.\ D {\bf 78}, 106005 (2008).
[arXiv:0808.1725 [hep-th]].
}
\lref\DanielssonGI{
  U.~H.~Danielsson and L.~Thorlacius,
  ``Black holes in asymptotically Lifshitz spacetime,''
JHEP {\bf 0903}, 070 (2009).
[arXiv:0812.5088 [hep-th]].
}
\lref\BertoldiVN{
  G.~Bertoldi, B.~A.~Burrington and A.~Peet,
  ``Black Holes in asymptotically Lifshitz spacetimes with arbitrary critical exponent,''
Phys.\ Rev.\ D {\bf 80}, 126003 (2009).
[arXiv:0905.3183 [hep-th]].
}


\lref\KrishnanPV{
  C.~Krishnan,
  ``Hidden Conformal Symmetries of Five-Dimensional Black Holes,''
JHEP {\bf 1007}, 039 (2010)
[arXiv:1004.3537 [hep-th]];
  D.~Chen, P.~Wang and  H.~Wu,
  ``Hidden conformal symmetry of rotating charged black holes,''
Gen.\ Rel.\ Grav.\  {\bf 43}, 181-190 (2011)
[arXiv:1005.1404 [gr-qc]];
  M.~Becker, S.~Cremonini and W.~Schulgin,
  ``Correlation Functions and Hidden Conformal Symmetry of Kerr Black Holes,''
JHEP {\bf 1009}, 022 (2010)
[arXiv:1005.3571 [hep-th]];
  H.~Wang, D.~Chen, B.~Mu and H.~Wu,
  ``Hidden conformal symmetry of extreme and non-extreme Einstein-Maxwell-Dilaton-Axion black holes,''
JHEP {\bf 1011}, 002 (2010)
[arXiv:1006.0439 [gr-qc]];
  C.~-M.~Chen, Y.~-M.~Huang, J.~-R.~Sun, M.~-F.~Wu and S.~-J.~Zou,
  ``On Holographic Dual of the Dyonic Reissner-Nordstrom Black Hole,''
Phys.\ Rev.\  {\bf D82}, 066003 (2010)
[arXiv:1006.4092 [hep-th]];
  I.~Agullo, J.~Navarro-Salas, G.~J.~Olmo and L.~Parker,
  ``Hawking radiation by Kerr black holes and conformal symmetry,''
Phys.\ Rev.\ Lett.\  {\bf 105}, 211305 (2010)
[arXiv:1006.4404 [hep-th]];
  K.~-N.~Shao and Z.~Zhang,
  ``Hidden Conformal Symmetry of Rotating Black Hole with four Charges,''
Phys.\ Rev.\  {\bf D83}, 106008 (2011)
[arXiv:1008.0585 [hep-th]];
  A.~M.~Ghezelbash, V.~Kamali and M.~R.~Setare,
  ``Hidden Conformal Symmetry of Kerr-Bolt Spacetimes,''
Phys.\ Rev.\  {\bf D82}, 124051 (2010)
[arXiv:1008.2189 [hep-th]];
  D.~A.~Lowe, I.~Messamah and A.~Skanata,
  ``Scaling dimensions in hidden Kerr/CFT,''
[arXiv:1105.2035 [hep-th]];
  S.~Bertini, S.~L.~Cacciatori and  D.~Klemm,
  ``Conformal structure of the Schwarzschild black hole,''
[arXiv:1106.0999 [hep-th]].
}
\lref\BertiniGA{
  S.~Bertini, S.~L.~Cacciatori, D.~Klemm,
  ``Conformal structure of the Schwarzschild black hole,''
[arXiv:1106.0999 [hep-th]].
}

\lref\MaldacenaDR{
  J.~M.~Maldacena and  L.~Maoz,
  ``Desingularization by rotation,''
JHEP {\bf 0212}, 055 (2002)
[hep-th/0012025].
}
\lref\LuninIZ{
  O.~Lunin, J.~M.~Maldacena and L.~Maoz,
  ``Gravity solutions for the D1-D5 system with angular momentum,''
[hep-th/0212210].
}
\lref\BalasubramanianRT{
  V.~Balasubramanian, J.~de Boer, E.~Keski-Vakkuri and S.~F.~Ross,
  ``Supersymmetric conical defects: Towards a string theoretic description of black hole formation,''
Phys.\ Rev.\  {\bf D64}, 064011 (2001)
[hep-th/0011217].
}
\lref\DijkgraafFQ{
  R.~Dijkgraaf, J.~M.~Maldacena, G.~W.~Moore and  E.~P.~Verlinde,
  ``A Black hole Farey tail,''
[hep-th/0005003].
}

\lref\CveticKVa{
  M.~Cveti\v c and  D.~Youm,
  ``Entropy of nonextreme charged rotating black holes in string theory,''
Phys.\ Rev.\  {\bf D54}, 2612-2620 (1996).
[hep-th/9603147].
}

\lref\CveticKV{
  M.~Cveti\v c and  D.~Youm,
  ``All the static spherically symmetric black holes of heterotic string on a six torus,''
Nucl.\ Phys.\  {\bf B472}, 249-267 (1996).
[hep-th/9512127].
}

\lref\CveticHP{
  M.~Cveti\v c and F.~Larsen,  ``Conformal Symmetry for General Black Holes,'' 
[arXiv: 1106.3341 [hep-th]], submitted to JHEP.
}

\lref\MaldacenaRE{
  J.~M.~Maldacena,
  ``The Large N limit of superconformal field theories and supergravity,''
Adv.\ Theor.\ Math.\ Phys.\ \ {\bf 2}, 231  (1998), [Int.\ J.\ Theor.\ Phys.\ \ {\bf 38}, 1113  (1999)].
[hep-th/9711200].
}

\lref\MaldacenaDE{
  J.~M.~Maldacena, A.~Strominger and E.~Witten,
  ``Black hole entropy in M theory,''
JHEP {\bf 9712}, 002 (1997).
[hep-th/9711053].
}

\lref\ChenKT{
  B.~Chen and J.~-j.~Zhang,
  ``General Hidden Conformal Symmetry of 4D Kerr-Newman and 5D Kerr Black Holes,''
JHEP {\bf 1108}, 114 (2011).
[arXiv:1107.0543 [hep-th]].
}
\lref\HuangYG{
  Y.~-C.~Huang and F.~-F.~Yuan,
  ``Hidden conformal symmetry of extremal Kaluza-Klein black hole in four dimensions,''
JHEP {\bf 1103}, 029 (2011).
[arXiv:1012.5453 [hep-th]].
}

\lref\ShaoCF{
  K.~-N.~Shao and Z.~Zhang,
  ``Hidden Conformal Symmetry of Rotating Black Hole with four Charges,''
Phys.\ Rev.\ D {\bf 83}, 106008 (2011).
[arXiv:1008.0585 [hep-th]].
}
\lref\ChenYU{
  C.~-M.~Chen, Y.~-M.~Huang, J.~-R.~Sun, M.~-F.~Wu and S.~-J.~Zou,
  ``On Holographic Dual of the Dyonic Reissner-Nordstrom Black Hole,''
Phys.\ Rev.\ D {\bf 82}, 066003 (2010).
[arXiv:1006.4092 [hep-th]].
}
\lref\ChenZWA{
  D.~Chen, P.~Wang and H.~Wu,
  ``Hidden conformal symmetry of rotating charged black holes,''
Gen.\ Rel.\ Grav.\  {\bf 43}, 181 (2011).
[arXiv:1005.1404 [gr-qc]].
}
\lref\ChenAS{
  C.~-M.~Chen and J.~-R.~Sun,
  ``Hidden Conformal Symmetry of the Reissner-Nordstr{\o}m Black Holes,''
JHEP {\bf 1008}, 034 (2010).
[arXiv:1004.3963 [hep-th]].
}

\lref\KastorGT{
  D.~Kastor and J.~H.~Traschen,
  ``A Very effective string model?,''
Phys.\ Rev.\ D {\bf 57}, 4862 (1998).
[hep-th/9707157].
}
\lref\CveticVP{
  M.~Cveti\v c and F.~Larsen,
  ``Black hole horizons and the thermodynamics of strings,''
Nucl.\ Phys.\ Proc.\ Suppl.\  {\bf 62}, 443 (1998), [Nucl.\ Phys.\ Proc.\ Suppl.\  {\bf 68}, 55 (1998)].
[hep-th/9708090].
}

\Title{\vbox{\baselineskip12pt 
\hbox{UPR-1234-T} 
\vskip-.5in}
}
{\vbox{\centerline{Conformal Symmetry for Black Holes in Four Dimensions}
}}
\medskip
\centerline{\it
Mirjam Cveti\v c${}^{1,2}$ and Finn Larsen${}^{2}$
}
\bigskip
\centerline{${}^1$Department of Physics and Astronomy, University of Pennsylvania, Philadelphia, PA-19104, USA.}
\centerline{${}^2$Center for Applied Mathematics and Theoretical Physics,
University of Maribor, Maribor, Slovenia.}\smallskip
\centerline{${}^3$Michigan Center for Theoretical Physics, 450 Church St., Ann Arbor,
MI-48109, USA.}
\smallskip

\vglue .3cm
\bigskip\bigskip\bigskip
\centerline{\bf Abstract}
\noindent
We show that the asymptotic boundary conditions of general asymptotically flat black holes in four dimensions can be modified such that a conformal symmetry emerges. The black holes with the 
asymptotic geometry removed in this manner satisfy the equations of motion of minimal supergravity
in five dimensions. We develop evidence that a two dimensional CFT dual of general black holes in four dimensions account for their black hole entropy. 

\Date{}

\newsec{Introduction}
For supersymmetric black holes it has long been understood how to account for the entropy of in terms of dual weakly coupled conformal field theories in two dimensions
(some reviews are \refs{\DavidWN,\KrausWN,\SenQY}). General arguments suggest that similar advances are precluded in settings that do not preserve supersymmetry even approximately. However, the general
 (non-supersymmetric) entropy formula takes a form that suggests a dual two dimensional conformal field theory even when supersymmetry is broken substantially \refs{\LarsenGE,\CveticKV}. Our recent investigation of this situation in the setting of five dimensional black holes lead to a concrete 
procedure that might 
account for the black hole entropy of black holes even in situations far from the supersymmetric limit \CveticHP. In our previous work we proposed that the entropy of 
general black boles in five dimensions can be addressed by an analysis with the 
following components: 

\indent
1. Establish that {\it the causal structure and the thermodynamics} of the black 
hole geometry is independent of a certain conformal factor. Accordingly this conformal factor can be interpreted as a specification of the {\it environment} of the black hole
in a manner decoupled from its internal structure. 

\indent 
2.
Show that {\it the scalar wave equation} exhibits $SL(2,R)\times SL(2,R)$ symmetry for 
some specifications of the conformal factor. We refer to the geometry with conformal factor modified in this manner as the ``subtracted'' geometry. The subtracted geometry has 
the same near horizon properties as the original black hole  but different asymptotics at large distances: it is not asymptotically flat.

\indent
3.
Show that {\it an auxiliary dimension} can be introduced that lifts the subtracted geometry 
to one dimension higher  such that both the separability and the $SL(2,R)\times SL(2,R)$ symmetry of the scalar wave equation become manifest. In this setting the 2D conformal symmetry is linearly realized by representing the subtracted geometry as a $U(1)$ coset.   
\medskip

In this paper we present computations that carry out these steps in the context of a large class of four dimensional black holes. This aspect of the present work is
an adaptation to four dimensions of the analogous computations in five dimensions previously presented in \CveticHP. This part of the paper generalizes the ``hidden conformal symmetry'' proposal \CastroFD\ to include charges and sharpens it by providing a formulation directly in the geometry rather than relying on the wave equation. It also greatly sharpens our own suggestions \CveticXV\ from over a decade ago. 

These generalizations are interesting in their own right. However, in addition this paper also addresses some of aspects of our approach that present legitimate questions:  

\indent 
1.
The subtracted black hole geometry does not generally satisfy the {\it equations of motion} since we simply change the geometry by declaration. We view this as acceptable, since the black hole ``itself'' cannot be in equilibrium unless it is surrounded by a ``box'' kept at the same temperature as the black hole. Such a box must be made from matter and it is usually not worthwhile to specify this matter explicitly. However, our construction is quite novel 
and its off-shell nature is one of its unusual features. In this paper we address this concern directly, by constructing in the non-rotating case  the explicit matter  such that the equations of motion are satisfied. 

\indent 
2. 
An appealing interpretation of the matter that supports the subtracted black hole 
solution in four dimensions is that it corresponds to {\it minimal} supergravity in five dimensions.

This paper is organized as follows.
In section 2 we introduce the general black holes in four dimensions and analyze their causal structure and their thermodynamics. We show that these features are independent of the conformal factor.  
In section 3 we analyze the scalar wave equation for general conformal factor. We determine the ``subtracted'' conformal factor such that the desirable features suggested by 
the standard conformal factor become exact. 
In section 4 we identify matter fields such that the black hole with subtracted conformal factor satisfies the equations of motion. 
In section 5 we exhibit the $SL(2,R)\times SL(2,R)$ symmetry of the subtracted geometry explicitly, and we use an auxiliary five dimensional geometry to argue that the symmetry is enhanced to the conformal group in two dimensions. 
In section 6 we discuss hidden conformal symmetry from a 4D and a 5D point of view. 

\newsec{General Black Holes in Four Dimensions}
In this section we introduce the black hole geometry in the fibered form we find useful. 
We review the causal structure of the black holes and derive their thermodynamics in a manner that exhibits independence of the conformal factor
$\Delta_0$.  

\subsec{The Black Hole Metric}
The setting for our discussion is the rotating black hole solution of four dimensional string theory with four independent $U(1)$ charges \CveticKV. The asymptotic charges of the 
black hole are parametrized as:
\eqn\baa{\eqalign{
G_4 M & = {1\over 4}m\sum_{I=0}^3\cosh 2\delta_I ~,\cr
G_4 Q_I & = {1\over 4}m\sinh 2\delta_I~,~(I=0,1,2,3)~,\cr
G_4 J & = m a (\Pi_c - \Pi_s)~,
}}
where we employ the abbreviations
\eqn\bab{
\Pi_c \equiv \prod_{I=0}^3\cosh\delta_I 
~,~~~ \Pi_s \equiv  \prod_{I=0}^3 \sinh\delta_I~.
}
The parametric mass and angular momentum $m, a$ both have dimension of length.  

We write the 4D metric as a fibration over a 3D base space
\eqn\ba{\eqalign{
ds^2_4 & = - \Delta^{-1/2}_0 G ( dt+{\cal A})^2 + \Delta^{1/2}_0 
\left( {dr^2\over X} + d\theta^2 + {X\over G} \sin^2\theta d\phi^2\right)~,
}}
where for the black holes we consider
\eqn\bb{\eqalign{
X & = r^2 - 2mr + a^2~, \cr
G & = r^2 - 2mr + a^2 \cos^2\theta ~, \cr
{\cal A} & = {2ma\sin^2\theta \over G}
\left[ (\Pi_c - \Pi_s) r + 2m\Pi_s\right] d\phi~,\cr
\Delta_0 &= \prod_{I=0}^3 (r + 2m\sinh^2 \delta_I)
+ 2 a^2 \cos^2\theta [r^2 + mr\sum_{I=0}^3\sinh^2\delta_I
+ 4m^2 (\Pi_c - \Pi_s)\Pi_s \cr & ~~~~~~- 2m^2 \sum_{I<J<K}
\sinh^2 \delta_I \sinh^2 \delta_J\sinh^2 \delta_K]
+ a^4 \cos^4\theta~.
}}
The fibered form \ba\ of the metric does not reduce to the one usually presented in textbooks for Kerr. However, the alternate form here simplifies manipulations significantly, especially 
when all the string theory charges are included. 

The rather complicated conformal factor $\Delta_0$ simplifies in some special cases. The benchmark is the non-rotating case $a=0$ where only the first term remains. However, the expression also simplifies with rotation when the four charges are equal in pairs
\eqn\bba{
\Delta_0 = [ (r+2m\sinh^2\delta_1)(r+2m\sinh^2\delta_2)+ a^2\cos^2\theta]^2~.
}
The generic case with rotation and four independent charges does not simplify. 

\subsec{Causal Structure}
It is instructive to analyze a general black hole geometry of the form \ba\ where 
$X$ is an arbitrary function of the radial variable $r$, the function $G$ is
\eqn\bbb{
G=X-a^2\sin^2\theta
~,}
and 
$\Delta_0, {\cal A}_\phi$ are arbitrary functions of both $r$ and the polar angle $\theta$. We will return to the specific forms \bb\ later. 

Trajectories along the Killing direction parametrized by $t$ cease to be 
time-like at the {\it static limit} where
\eqn\ca{
G=0~.
}
The volume inside this surface (but outside the event horizon) is the
{\it ergosphere}.

In the ergosphere physical trajectories at fixed $r,\theta$ have a nontrivial component along the azimuthal angle $\phi$ because they would be spacelike if they were fully directed along $t$. However, {\it all} directions at fixed $r,\theta$ become space-like once the determinant in the $t-\phi$ plane 
\eqn\cb{
{\rm det} (t-\phi) = - X\sin^2\theta~,
}
turns positive. This identifies the event horizon as the surface
\eqn\cc{
X=0~.
}
The relation \bbb\ ensures that the event horizon \cc\ is indeed inside the static limit \ca, except at the poles $\sin\theta=0$ where the two surfaces meet. 

\subsec{Black Hole Thermodynamics}
It is worthwhile to analyze the thermodynamics of the black holes while remaining in the general setting with $X, \Delta_0, {\cal A}_\phi$ unspecified and $G$ given by \bbb. 

We first present the metric \ba\ as 
\eqn\cd{
ds^2_4 = \Delta^{1/2}_0 \left({dr^2\over X} + X {\sin^2\theta\over G} d\phi^2\right)
+[\Delta_0^{1/2} d\theta^2 - \Delta^{-1/2}_0 G (dt+{\cal A}_\phi d\phi)^2]~,
}
and then focus on the region $X\sim 0$ near the event horizon \cc. In this region 
$G\sim -a^2\sin^2\theta<0$ so the geometry in the round bracket is independent 
of $\theta$. Moreover, this part of the geometry has Lorentzian signature and can be interpreted as Rindler space\foot{We assume the horizon $X=0$ is a simple pole in $X(r)$. A double pole corresponds to an extremal black holes which presents a challenge for a thermodynamic interpretation, as usual.}.
We present the acceleration of this Rindler space in terms of the Euclidean period
\eqn\ce{
\beta_\phi = {4\pi a\over\left(\partial_r X\right)_{\rm hor}}~,
} 
determined such that the ``time'' $\phi$ avoids a conical singularity. 

The geometry of the black hole horizon is encoded in the square bracket  of \cd. 
Recalling again that $G\sim -a^2\sin^2\theta<0$ it is recognized that the horizon has topology $S^2$ as expected. 
The determination of the Euclidean period \ce\ was carried out at 
any point on the event horizon so the geometry in 
the square bracket must be kept fixed as $\phi$ is periodically identified. 
This consideration determines the Euclidean periodicity of the asymptotic time $t$ as 
\eqn\cf{
\beta_H = - \left({\cal A}_\phi\right)_{\rm hor} \beta_\phi~.
}
We can interpret this formula in Lorentzian signature where it gives the rotational 
velocity
\eqn\cfa{
\Omega_H = {\beta_\phi\over\beta_H} = - {1\over ({\cal A_\phi})_{\rm hor}}~.
}

We will find it useful to introduce the {\it reduced} angular potential that has some of the overall 
factors removed
\eqn\cfz{
{\cal A}_{\rm red} = {G\over a\sin^2\theta}{\cal A}_\phi \sim -a ({\cal A}_\phi)_{\rm hor}~.
}
The explicit solution \bb\ gives a reduced angular potential ${\cal A}_{\rm red}$
that only depends on the radial coordinate. This in turn ensures that the 
rotational velocity \cfa\ is independent of the position on the horizon, as it should be. 
This property motivates the use of the reduced angular potential ${\cal A}_{\rm red}$ 
without reference to the explicit solution. 

With the notation \cfz\ and the general result \ce\ we can rewrite the inverse temperature
\cf\ as 
\eqn\cfaa{
\beta_H  = \left({4\pi {\cal A}_{\rm red}\over \partial_r X}\right)_{\rm hor}~.
}
This expression is manifestly independent of the polar angle. 

The black hole entropy is computed from the area of the event horizon. 
Reading the measure from the square bracket of \cd\ (at fixed asymptotic 
time $t$) we find
\eqn\cfb{
S = {1\over 4 G_4} \int_{r=r_+} \sqrt{- G{\cal A}^2_\phi} d\phi d\theta
= {a\over 4 G_4} \int_{r=r_+} | {\cal A}_\phi | \sin\theta d\phi d\theta
= {\pi({\cal A}_{\rm red})_{\rm hor}\over G_4}~.
}
The integral was evaluated by exploiting the constancy of the rotational velocity \cfa\ on the event horizon. 

The formula \cfb\ identifies the black hole entropy with the reduced potential 
${\cal A}_{\rm red}$, up to a universal constant. This is surprising because the 
introduction of the potential ${\cal A}_{\rm red}$ in \cfz\ relies on rotation of the black hole. The result nevertheless makes sense because the actual value or ${\cal A}_{\rm red}$ on 
the horizon remains finite in the limit of vanishing angular momentum. More generally, our derivation of the inverse temperature \cfaa\ and the black hole entropy \cfb\ relies on the rotation at intermediate steps but the final expressions are finite in the non-rotating limit and in agreement with those obtained using other methods. We will see angular momentum in a privileged role repeatedly in this work. 

An important corollary to the expressions \ce, \cfa, \cfaa, \cfb\ for the
thermodynamic parameters is their independence of the conformal 
factor $\Delta_0$. We interpret this to mean that $\Delta_0$ characterizes
the environment of the black hole rather than its internal structure.

\subsec{Thermodynamics: Explicit Expressions}
In the case of the explicit function $X$ given in \bb, the event horizon is at the largest solution to the quadratic equation \cc\
\eqn\cg{
r_+ = m + \sqrt{m^2 - a^2}~.
}
For the solutions \bb\ the thermodynamic potential \ce\ becomes
\eqn\cj{
\beta_H\Omega =\beta_\phi= {2\pi a\over r_+ - m } = {2\pi a \over\sqrt{m^2 - a^2}}~,
}
and the reduced potential \cfz\ is
\eqn\daa{
{\cal A}_{\rm red} = 2m [ (\Pi_c - \Pi_s)r + 2m \Pi_s]~.
}
Then the inverse Hawking temperature \cfaa\ yields
\eqn\ch{
\beta_H = {4\pi m\over r_+ - m }[ (\Pi_c - \Pi_s)r_+ + 2m\Pi_s]
= 4\pi m\left(  {m\over\sqrt{m^2-a^2}}(\Pi_c + \Pi_s) + (\Pi_c - \Pi_s)\right)~,
}
and the black hole entropy \cfb\ becomes 
\eqn\ck{\eqalign{
S &= {2\pi m\over G_4} [ (\Pi_c - \Pi_s)r_+ + 2m \Pi_s] = {2\pi m\over G_4}\left( (\Pi_c + \Pi_s)m + (\Pi_c - \Pi_s)\sqrt{m^2-a^2} \right)~.
}}
The thermodynamic potentials \cj, \ch, and \ck\ agree with those found in \CveticKVa\ using conventional methods.

\newsec{The Subtracted Geometry}
The massless wave equation hints at a dual 2D CFT even for the general black holes we consider. This section discusses the appearance of hypergeometric structure, the precursor of $SL(2,R)\times SL(2,R)$ symmetry.

\subsec{Separability of the Scalar Wave Equation}
As in the previous section it is instructive to first consider a general metric of 
the form \ba\ where $X$, $G, \Delta_0, {\cal A}_\phi$ are arbitrary functions, except for 
the rudimentary assumptions stated around \bbb. We will also assume that ${\cal A}_{\rm red}$ introduced in \cfz\ depends on $r$ alone, as it does in our primary example. 

The Laplacian derived by inverting the metric \ba\ becomes  
\eqn\da{
\Delta^{-1/2}_0\left[
\partial_r X \partial_r + {1\over\sin\theta} \partial_\theta
\sin\theta \partial_\theta - {\Delta_0\over G}\partial^2_t
+{G\over X\sin^2\theta} (\partial_\phi - {\cal A}_\phi\partial_t)^2
\right]~.
}
The last two terms generally mix $r$ and $\theta$ in a complicated way that obstructs
separability. At this point we utililize \bbb\ for $G$ and also introduce ${\cal A}_{\rm red}$ from \cfz. Then the Laplacian simplifies to
\eqn\db{
\Delta^{-1/2}_0\left[
\partial_r X \partial_r - {1\over X} ( {\cal A}_{\rm red} \partial_t + a\partial_\phi)^2+ {1\over\sin\theta} \partial_\theta
\sin\theta \partial_\theta + {1\over\sin^2\theta} \partial^2_\phi
+ {  {\cal A}^2_{\rm red}- \Delta_0\over G}\partial^2_t
\right]~.
}
In this equation it is just the last term that prevents separability, when disregarding 
(for now\foot{We will later identify a non-minimal coupling that ensures separability of massive scalars as well.}) the overall factor of $\Delta^{-1/2}_0$.

In the actual geometry \bb\ the reduced potential ${\cal A}_{\rm red}$ is \daa\ and the combination
$\Delta_0- {\cal A}^2_{\rm red}$ contains a factor $G$ that ensures factorization
\eqn\dc{
{ \Delta_0- {\cal A}^2_{\rm red} \over G} = r^2 + 2mr\left( 1+\sum_{I=0}^3s^2_I\right) + 8m^2 (\Pi_c-\Pi_s)\Pi_s  - 4m^2 \sum_{I<J<K}
s^2_I s^2_J s^2_K
+ a^2 \cos^2\theta~,
}
where $s^2_i \equiv \sinh^2\delta_i$. This expression implies separability of the Laplacian \db\ and so separability of the (massless) wave equation. The details of the expression
still looks quite forbidding but this is primarily due to an intricate dependence on black hole parameters. The right hand side is in fact just a quadratic polynomial in $r$, with the constant term depending quadratically on $\cos\theta$.

\subsec{The Subtracted Geometry}
The differential equation \db\ with the effective potential \dc\ has simplifying features beyond the seperability. The radial 
equation has two {\it regular} singularities, at $r=r_+$ and $r=r_-$. The indices at these regular singularities are ${i\beta_\pm\over 2\pi}$, ie. essentially the outer and inner horizon temperatures. The radial equation has a third singularity at infinity, but this singularity is 
irregular. If it had been regular the radial equation would have been  the hypergeometric equation, with its $SL(2,R)$ symmetry permuting the three singularities. This situation is desirable because it would hint at an underlying conformal symmetry. 

The irregular singularity at infinity is due to the asymptotic behavior $\Delta_0\sim r^4$ 
at large $r$ which encodes the asymptotic flatness of spacetime. If instead $\Delta_0\to \Delta\sim r^2$ the singularity at infinity in the radial equation would be regular. For even more 
special warp factors with $\Delta_0\to \Delta\sim r$ at large $r$ the radial equation maintains its hypergeometric character but, in addition, the angular equation simplifies to the familiar spherically symmetric form 
\eqn\dh{
\left(  {1\over\sin\theta} \partial_\theta
\sin\theta \partial_\theta + {1\over\sin^2\theta} \partial^2_\phi\right) \chi(\theta,\phi)
 = - l(l+1)\chi(\theta,\phi)~.
}
When $\Delta\sim r$ the geometry thus indicates an unbroken $SU(2)$ R-symmetry. 
In this case the indices of the regular singularity at infinity are $(l, -l -1)$. 

It was shown in section 2 that both the causal structure and the thermodynamics of black holes is independent of the conformal factor $\Delta_0$. We interpret this as a demonstration that an alternate $\Delta_0\to \Delta$ corresponds to a black hole with the same internal structure as the original one, but a black hole that finds itself in a different external environment. 

We will focus on  the warp factors $\Delta$ that preserve separability of the scalar wave equation and also analyticity in the coordinates. These technical assumptions
identify a warp factor with the asymptotic behavior $\Delta\sim r$ uniquely as
\eqn\de{\eqalign{
\Delta_0\to \Delta 
&= (2m)^3 r (\Pi^2_c - \Pi^2_s) + (2m)^4 \Pi^2_s - (2m)^2 (\Pi_c-\Pi_s)^2 a^2\cos^2\theta~.
}}
In particular, the condition of separability determines the $\theta$-dependence by the requirement that $\Delta-{\cal A}_{\rm red}^2$ be factorizable by $G$:
\eqn\deg{
{\Delta-{\cal A}_{\rm red}^2\over G} = - 4m^2 ( \Pi_c-\Pi_s)^2~.
}
The condition of separability is powerful even in the nonrotating limit $a\to 0$. For example, the analysis of the Schwarzchild geometry geometry presented in \BertiniGA\ is not consistent with this criterion. 

%
%
%

More general assignments of the conformal factor are possible if we allow
asymptotic behavior $\Delta\sim r^2$ while maintaining separability. Such alternate conformal factors take the form $\Delta' = {\rm const} \cdot G + \Delta$ where $\Delta$ is given in \deg.  The angular equation of such geometries generalize \dh\ in a manner that, for the purpose of the radial equation, can be absorbed in a renormalization of the separation constant $l\to l_{\rm eff}$. The resulting effective angular momenta $l_{\rm eff}$ generally differ from the non-negative integers assigned to $l$. Indeed, they do not even have to be real: complex $l_{\rm eff}$ are interpreted physically in terms of superradiance, in the case of near extreme Kerr \refs{\BredbergPV,\CveticJN}. In this paper we primarily analyze the minimally subtracted metric with conformal factor \dh. 

As we have explained, the motivation for changing the conformal factor $\Delta$ at will is the interpretation that such changes do nothing to alter the internal structure of the black hole, it  merely changes the environment of the quantum black hole. This argument is not beyond reproach. It is therefore worth noting a less ambitious reasoning that motivates the same procedure: {\it any} setting where the scalar wave equation approximates the hypergeometric equation can be interpreted as a situation where the actual conformal factor $\Delta_0$ is well approximated by the $\Delta$ that gives the hypergeometric equation exactly. In any such setting we might as well consider the approximate geometry from the outset. \foot{In the special case of a dilute gas (near-BPS)  approximation where $\delta_I>>1$, ($I=1,2,3$), the exact warp factor $\Delta_0$ and the subtracted one $\Delta$ coincide, as expected.} The decoupling limit of the AdS/CFT correspondence is a special case of this reasoning, the near extreme limit of Kerr is another. In such settings the present point of view offers the advantage that we are not limited to the scalar wave equation. The geometry with enhanced symmetry allows the analysis of many other questions. 
 
 \subsec{Kerr}
It is worthwhile making our considerations more explicit in the case of the pure Kerr black hole, without any charges. In this case the parametric mass and angular momenta are 
$m=G_4 M$ and $a= J/M$. The full conformal factor in \bb\ (or \bba) becomes
\eqn\dga{
\Delta_0 = (r^2  +a^2 \cos^2\theta)^2~.
}
Separability for pure Kerr can be traced to factorization of the effective potential
\eqn\dgc{
{\Delta_0- {\cal A}^2_{\rm red}\over G} = r^2 + 2mr + a^2\cos^2\theta~,
}
with ${\cal A}_{\rm red}=2mr$. 

The conformal factor \de\ in the subtracted geometry simplifies to 
\eqn\dgb{
\Delta_0\to \Delta = 4m^2 ( 2mr -a^2\cos^2\theta)~.
}
The subtracted geometry still has ${\cal A}_{\rm red}=2mr$ and separability now 
relies on the factorization
\eqn\dgd{
{\Delta- {\cal A}^2_{\rm red}\over G} = -4m^2~.
}
The additional hypergeometric structure is due to the improved asymptotic behavior 
of the effective potential. 

The special case of near-extreme Kerr can be usefully analyzed by focussing on the near horizon region isolated by the NHEK limit \BardeenPX. The NHEK scaling limit tunes the parameters of the 
black hole so $\sqrt{m^2-a^2}=\epsilon\lambda\to 0$ as $\lambda\to 0$ while simultaneously focussing on the near horizon region where $r-m=\lambda U\to 0$. 
In the NHEK limit the black hole reduces to a warped AdS$_3$ geometry with radius $\ell^2=2m^2$ and warp factor $\Omega={1\over 2}(1+\cos^2\theta)$. 
 
Applying the NHEK limit directly on the conformal factor \dga\ we recover the NHEK warp factor
\eqn\dge{
\sqrt{\Delta_0}\to \ell^2 {1\over 2}(1+\cos^2\theta)~,
}
If we instead apply the NHEK limit on the subtracted conformal factor \dgb\ we
find the warp factor
\eqn\dgf{
\sqrt{\Delta}\to \ell^2 \sqrt{1 + \sin^2\theta}~.
}
It is clear from this comparison that our subtraction procedure differs from the NHEK 
limit even for the rapidly spinning black holes where both analyses apply.  In either approach the scalar field equation is in hypergeometric form and it is possible that this means there are two valid CFT descriptions for rapidly spinning black holes.
 It would of course be interesting to find a relation between these descriptions. 
This in no way presents a contradiction: both of these schemes involve
 a scalar field equation of a hypergeometric form and it is possible that this means there are two valid CFT descriptions for rapidly spinning black holes. It would of course be interesting to find a relation between these descriptions.

\subsec{Asymptotic Behavior of the Subtracted Geometry}
The asymptotic behavior of the subtracted geometry is 
\eqn\dft{
ds^2_4 \sim - \Delta^{-1/2} r^2 dt^2 + \Delta^{1/2} {1\over r^2} (dr^2 + r^2 d\Omega_2^2)~,
 }
where $\Delta\sim \ell^3 r$ with $\ell^3= (2m)^3(\Pi^2_c - \Pi^2_s)$. 
We can introduce $R=4\ell^{3/4}r^{1/4}$ and write this asymptotic behavior as
\eqn\dgt{
ds^2_4 \sim - \left({R\over 4\ell}\right)^6 dt^2 + dR^2 + \left({R\over 4}\right)^2 d\Omega_2^2~.
}
In this form the asymptotic geometry has an obvious scaling symmetry 
$ds^2_4\to \lambda^2 s^2_4$ that is implemented by taking 
$R\to\lambda R$ and $t\to\lambda^{-2}t$. The nonstandard scaling of time 
is reminiscent of the Lifshitz symmetry that has recently been developed for
applications of holography to condensed matter systems (some representative works are
\refs{\KachruYH,\DanielssonGI,\BertoldiVN}).
It would be interesting to develop the asymptotic symmetry of the present 
context in more detail.

\newsec{The Matter Supporting the Geometry}
The black hole metric with subtracted conformal factor does not {\it a priori} satisfy 
the equations of motion. In this section we explicitly identify a matter configuration 
that supports the geometry. 

\subsec{Physical Matter}
Our reasoning up to this stage has been to freely modify the conformal factor 
$\Delta_0\to\Delta$ as needed in order that the scalar wave equation exhibits enhanced symmetries, without the black hole thermodynamics and causal structure having been modified. This lead us to the specific assignment \de\ for the conformal factor, while keeping the remaining parts of the solution \bb\ intact. 

The subtracted geometry with conformal factor $\Delta$ does not satisfy the 
equations of motion with the matter that was specified before the subtraction 
procedure. In particular, the Kerr geometry with subtracted conformal factor is not 
a vacuum solution. However, we can form a genuine solution by specifying appropriate matter that supports the solution. 

To assess the situation we focus for now on the non-rotating solutions. The 
Einstein tensor for the geometries \bb\ with arbitrary conformal 
factor $\Delta=e^{-4U}$ then becomes
\eqn\fa{\eqalign{
G_{{\hat\phi}{\hat\phi}} & = G_{{\hat\theta}{\hat\theta}} = -G_{{\hat r}{\hat r}}
= e^{2U} \big(r\partial_r U + 1\big) \big( (r-2m)\partial_r U + 1\big)~,\cr
G_{{\hat t}{\hat t}} & = G_{{\hat\theta}{\hat\theta}}  + 2e^{2U}r(r-2m)
\big(\partial_r^2 U - (\partial_r U)^2\big)~.
}}
The hatted coordinates refer to the standard orthonormal frame. We wish to find matter with an energy-momentum tensor that equates this Einstein tensor. 

It turns out that for the static case it is sufficient to consider the STU Lagrangian in the absence of pseudoscalars:
\eqn\fb{
{\cal L} = -{1\over 16\pi G_4} ( R -
{1\over 2}\sum_{i=1}^3\nabla_\mu \eta_i \nabla^\mu \eta_i - {1\over 4}
e^{-\eta_1 - \eta_2 -\eta_3} F^0_{\mu\nu}F^{0\mu\nu}-{1\over 4}e^{-\eta_1 -\eta_2-\eta_3}\sum_{i=1}^3 e^{2\eta_i}
F^i_{\mu\nu}F^{i\mu\nu}).
}
In the case of a spherically symmetric configuration (without magnetic fields) 
the energy momentum becomes
\eqn\fc{\eqalign{
8\pi G_4 T_{{\hat\phi}{\hat\phi}} & = 8\pi G_4 T_{{\hat\theta}{\hat\theta}} = 
-8\pi G_4 T_{{\hat r}{\hat r}}~,\cr
8\pi G_4 T_{{\hat\phi}{\hat\phi}} & = {1\over 4} Xe^{2U} \sum_{i=1}^3(\partial_r\eta_i)^2   - {1\over 4}e^{-\eta_1 - \eta_2 -\eta_3} 
\left((F^0_{rt})^2+
\sum_{i=1}^3 e^{2\eta_i}(F^i_{rt})^2 \right)~,
\cr
8\pi G_4 T_{{\hat t}{\hat t}} &  = {1\over 4} Xe^{2U}\sum_{i=1}^3 (\partial_r\eta_i)^2 + {1\over 4}e^{-\eta_1 - \eta_2 -\eta_3}
\left( (F^0_{rt})^2+\sum_{i=1}^3 e^{2\eta_i}
(F^i_{rt})^2 \right)~.
}}
As a check on the equations we may compute the Einstein tensor \fa\ for 
the standard conformal factor 
$\Delta_0=e^{-4U}=\prod_{I=0}^3 h_I$ with $h_I = r + 2m\sinh^2\delta_I$ and
verify that the result agrees with the energy-momentum tensor \fc\ for the matter 
\eqn\fd{\eqalign{
e^{-\eta_i} &= h_i\sqrt{h^0\over h_1 h_2 h_3} ~,~i=1,2,3~,\cr
A^I_t & = {2m\sinh\delta_I \cosh\delta_I\over h_I}~,~I=0,1,2,3~.
}}
This is the matter that supports the solution with standard conformal factor. 

The desired Einstein tensor \fa\ satisfies $G_{{\hat\phi}{\hat\phi}} = G_{{\hat\theta}{\hat\theta}} = -G_{{\hat r}{\hat r}}$ even for an arbitrary conformal factor. 
Comparing with the energy momentum tensor \fc\ for the STU-model we see that a combination of scalars and vectors will be appropriate matter also in the general case. 

The specific conformal factor \de\ that we focus on corresponds to
\eqn\fe{
U = -{1\over 4} \ln \left( (2m)^3 ( r (\Pi_c^2- \Pi_s^2) + 2m\Pi_s^2)\right) ~.
}
Evaluating the right hand side of \fa\ for this $U$ we determine the scalar and vector matter needed to support the subtracted solution as:
\eqn\ff{\eqalign{
8\pi G_4 T^{\rm scalar}_{{\hat t}{\hat t}} &  = {1\over 2}(G_{{\hat t}{\hat t}} - G_{{\hat\theta}{\hat\theta}}) ={3\over 16}Xe^{2U} 
{(\Pi_c^2- \Pi_s^2)^2\over ( r (\Pi_c^2- \Pi_s^2) + 2m\Pi_s^2)^2}~,\cr
8\pi G_4 T^{\rm vector}_{{\hat t}{\hat t}} & = {1\over 2}(G_{{\hat t}{\hat t}} + G_{{\hat\theta}{\hat\theta}}) =
e^{2U} \left[ {3\over 4} + {(2m)^2\Pi_s^2\Pi_c^2\over 4( r (\Pi_c^2- \Pi_s^2) + 2m\Pi_s^2)^2}
\right]~.
}}
At this point comparison with the scalar and vector terms in \fc\ gives simple ordinary differential equations for the matter. 

The solution for the matter is not unique. For example, a duality transformation will leave the Einstein geometry invariant but change the matter. We will construct just the simplest solution and note just one obvious ambiguity.
Equating the first line in \ff\ with the scalar term in \fc\ we find (for some choice of integration constant):
\eqn\fg{
\eta_i = - {1\over 2} \ln \left( (2m)^3r(\Pi_c^2- \Pi_s^2) + \Pi_s^2\right)~~,~i=1,2,3~.
}
\eqn\fh{\eqalign{
F^i_{tr} &= e^{-{1\over 2}\eta_i+U} = 1~~,~i=1,2,3~,\cr
F^0_{tr} & = {\Pi_c \Pi_s \over 4m^2(r (\Pi^2_c -\Pi_s^2) + 2m\Pi_s^2)^2}~.
}}
A different integration constant in \fg\ would rescale $F^i_{rt}$ by some factor  $e^{-{1\over 2}\delta\eta}$
(with $\delta\eta$ some constant) and simultaneously rescale $F^0_{rt}$ by $ e^{{3\over 2}\delta\eta}$. 
For example, the addition of $2\ln 2m$ to the right hand side of \fg\ makes the scalar 
field $\eta_i$ dimensionless and gives $F^i$ and $F^0$ the same dimension (of 
inverse length). The choice made in \fg\ avoids the introduction of an arbitrary scale (like $2m$) and will be convenient later. 

The scalar and vector fields \fg,\fh\ were constructed such that the Einstein equations are satisfied. It remains to verify that the matter field equations of motion are also obeyed. 
This is a straightforward exercise. 

\subsec{Discussion of Matter}
{\it Any} geometry is a solution to Einstein's equations, if the energy momentum tensor is chosen as the Einstein tensor of the geometry. The nontrivial question is always whether such matter is physical. The standard criterion is to ask whether the matter specified by the Einstein tensor satisfies suitable energy conditions.
In the present situation {\it all} of the standard energy conditions 
--- dominant,  strong, weak, and null --- are in fact satisfied (at least when there is no rotation). 
The black holes with subtracted conformal factor are therefore physical.

However, it is significant that we have gone beyond these criteria, by finding explicit 
matter. Indeed, we have shown that the subtracted geometries are solutions to the 
same theory as the original asymptotically flat black holes. This is 
explicit evidence that the black holes with subtracted conformal factor are physical. 
But it is presumably also useful for analyzing the physics of these solutions. 

The realization of the subtracted geometries as solutions to the same theory as the original black holes suggests that there is a more direct relation between these geometries. We expect that it can be obtained by a solution generating technique  
(within STU-model) on the original black hole. Actually, in the Schwarzschild case the subtracted geometry emerges as a specific Harrison transformation within Dilaton-Maxwell-Einstein gravity,  somewhat akin to transformations considered  within 
Maxwell-Einstein gravity  in \BertiniGA.  The identification of a transformation relating the true matter to the auxiliary matter \fg, \fh\ that we have identified in the non-rotating case could also serve as a practical strategy for generalizations to the cases 
with rotation. It would be interesting to compare the result of such considerations with the matter inferred from another method in the next section. 

\newsec{A Five Dimensional Interpretation of the Subtracted Geometry}
In this section we realize separability of the scalar wave equation geometrically,
by introducing an auxiliary dimension. The construction identifies a locally 
AdS$_3$ geometry that accounts for the $SL(2,R)\times SL(2,R)$ symmetry of the hypergeometric radial equation. It also gives a simplified representation of the matter 
that supports the subtracted geometry. 

\subsec{Simplified Derivation of the Scalar Wave Equation}
A useful initial goal is to seek a geometric interpretation of separability by attempting to construct a factorized spacetime directly. One of the ways that the geometry \ba\ couples the angular and radial coordinates is through the function $G$. It is therefore advantageous to expand the geometry such that the spurious poles at $G=0$ cancel explicitly:
\eqn\ea{\eqalign{
\Delta^{-1/2} ds^2_4 & = - {G\over\Delta} (dt+{a\sin^2\theta\over G}{\cal A}_{\rm red}d\phi)^2 + {X\sin^2\theta\over G}d\phi^2 + {dr^2\over X} + d\theta^2
\cr 
& = {1\over 4m^2 (\Pi_c-\Pi_s)^2} dt^2  - {1\over 4m^2 (\Pi_c-\Pi_s)^2 \Delta} [ {\cal A}_{\rm red} dt + 4m^2(\Pi_c-\Pi_s)^2 a\sin^2\theta d\phi]^2
 \cr & + {dr^2 \over X}+ d\theta^2 + \sin^2\theta d\phi^2~.
}}
We used \bbb\ relating $G$ to $X$ and \deg\ relating $\Delta$ and ${\cal A}_{\rm red}^2$. 
We can disentangle radial and polar variables further by writing the conformal 
factor \de\ as $\Delta=\rho+\gamma$ where
\eqn\eb{\eqalign{
\rho &= {\cal A}_{\rm red}^2 - 4m^2 (\Pi_c - \Pi_s)^2 X
= 8m^3 [ r  (\Pi_c^2 - \Pi_s^2)+ 2m \Pi^2_s - {a^2\over 2m}(\Pi_c - \Pi_s)^2  ]~,\cr
\gamma &= 4m^2 (\Pi_c - \Pi_s)^2 a^2 \sin^2\theta~.
}}
The subtracted metric \ea\ now simplifies to
\eqn\ed{\eqalign{
 ds^2_4 & = \Delta^{1/2}\left( - {X\over\rho} dt^2 + {dr^2 \over X}+ d\theta^2\right)+ \Delta^{-1/2} \rho\sin^2\theta 
(d\phi - {a{\cal A}_{\rm red}\over\rho}dt)^2~.
}}
We used the identity
\eqn\ec{
{\rho\gamma\over\rho+\gamma} \left( {p\over \gamma} - {q\over\rho}\right)^2 = 
{p^2\over\gamma}
+ {q^2 \over\rho} - { (p+q)^2\over\gamma+\rho}~,
}
with the identifications
\eqn\ebh{\eqalign{
p & = 4m^2 (\Pi_c - \Pi_s)^2 a\sin^2\theta d\phi~,\cr
q & = {\cal A}_{\rm red}dt~.
}}

The metric in the form \ed\ almost decouple the angular and radial dependence. For example, we can use this expression to separate variables in the Laplacian quite easily:
\eqn\eda{\eqalign{
&{1\over\sqrt{-g}}\partial_\mu\left(\sqrt{-g}g^{\mu\nu}\right)\partial_\nu\cr
& = \Delta^{-1/2} \left[ - {\rho\over X}\left(\partial_t + {a{\cal A}_{\rm red}\over\rho}\partial_\phi\right)^2
+ {1\over\sin\theta}\partial_\theta\sin\theta\partial_\theta + X\partial^2_r
+{\gamma+\rho\over\rho\sin^2\theta}\partial_\phi^2\right]  \cr
& = \Delta^{-1/2} \left[ - {1\over X}\left(\rho\partial^2_t 
+ 2a{\cal A}_{\rm red}\partial_\phi\partial_t + a^2\partial^2_\phi\right)
 + X\partial^2_r+ {1\over\sin\theta}\partial_\theta\sin\theta\partial_\theta
+ {1\over\sin^2\theta}\partial_\phi^2 \right]~.
}}
The recovery of the correct scalar wave equation gives a check on our algebra. 

\subsec{A Five Dimensional Lift}
It is instructive to reconsider separability from a five dimensional point of view. 
The last form of the metric in \ea\ is a good starting point. In that expression the second 
term is quite awkward but it can be presented as $-\Delta{\cal B}^2$ 
where
\eqn\eg{
{\cal B} = {p+q\over 2m(\Pi_c - \Pi_s)\Delta} = {
\left( (\Pi_c - \Pi_s)r + 2m\Pi_s\right) dt+ 2m(\Pi_c - \Pi_s)^2 a\sin^2\theta d\phi\over (\Pi_c - \Pi_s)\Delta }~.
}
It is natural to cancel this term by introducing an auxiliary coordinate $\alpha$
and so consider the  five dimensional auxiliary metric 
\eqn\ef{\eqalign{
&ds^2_5 = \Delta ( d\alpha + {\cal B})^2 + \Delta^{-1/2} ds^2_4 \cr
 &=  - {X\over\rho}dt^2 +{dr^2\over X} +  \rho
(d\alpha + { {\cal A}_{\rm red}\over 2m (\Pi_c - \Pi_s)\rho}dt)^2 + d\theta^2 + \sin^2\theta(d\phi + 2ma(\Pi_c - \Pi_s)d\alpha)^2.
}}
This geometry is locally AdS$_3\times S^2$. The sphere is fibered over the AdS$_3$
base by the shifted angle $\phi' = \phi + 2ma( \Pi_c - \Pi_s)\alpha = \phi + 2G_4J\alpha$. 
This does not prevent the product form from making separability explicit. Additionally, the
AdS$_3$ accounts for the hypergeometric form of the scalar wave equation.   Furthermore, this five dimensional lift 
 has the same geometry as the one obtained in the dilute gas approximation $\delta_I\gg 1$ ($I=1,2,3$) \CveticJA .

The auxiliary five dimensional geometry \ef\ was introduced as a dimensionless geometry, without a specific scale. Therefore the radius of the sphere $\ell_S=1$ is a pure number. Similarly the scale $\ell_A=2$ of the AdS$_3$ factor is a pure number. A related issue 
is that the auxiliary coordinate $\alpha$ has dimension length$^{-2}$. Assuming that
$\alpha$ is periodic with periodicity $2\pi R_\alpha$, the radius $R_\alpha$ will
have dimension of length$^{-2}$ as well. It is preferable to keep these awkward assignments of dimensions rather than introducing a specific scale that would 
in any case be arbitrary. 

An additional benefit of the five dimensional representation of the black hole is that it provides a geometrical interpretation of the matter supporting the subtracted solution, previously introduced for the non-rotating case in \fg, \fh. To see this we compute
the electric field strength for the gauge field ${\cal B}$ \eg\ in the non-rotating case
\eqn\efu{
F^{\cal B}_{tr} = {\Pi_c \Pi_s\over 4m^2((\Pi_c^2 - \Pi_s^2)r + 2m\Pi_s^2)^2}~.
}
This expression is identical to $F^0_{tr}$ given in \fh. We can therefore identify the gauge field in 4D with the graviphoton gauge field 
\eg, at least in the non-rotating case. The scalar field \fg\ 
introduced directly in four dimensions can similarly be identified with the dilaton determined through
\eqn\eft{
e^{-2\Phi_4}= R_\alpha\sqrt{\Delta}~.
}
The precise identification is $2\Phi_4=\eta_i~(i=1,2,3)$. The remaining gauge fields $F^i_{tr}$ introduced in \fh\ are constants that can be identified with a constant 
field strength in five dimensions. Thus the overall representation is  the one where the subtracted black hole geometry is a solution to {\it minimal} supergravity in five dimensions.  

\subsec{The Effective BTZ Black Hole}
It is worthwhile to rewrite the five dimensional auxiliary metric \ef\ explicitly 
as BTZ$\times S^2$ with the BTZ black hole presented in the standard form
\eqn\efd{
ds^2_{\rm BTZ} = - {(\rho^2_3-\rho^2_+)(\rho^2_3-\rho^2_-)\over\ell^2_A\rho^2_3}
dt^2_3 + {\ell^2_A\rho^2_3\over(\rho^2_3-\rho^2_+)(\rho^2_3-\rho^2_-)}d\rho^2_3
+\rho^2_3 (d\phi_3 + {\rho_+\rho_-\over\ell_A\rho^2_3}dt_3)^2~.
}
The BTZ coordinates are identified as
\eqn\efe{\eqalign{
\rho^2_3 & = (2mR_\alpha)^2 [ 2mr(\Pi^2_c - \Pi^2_s) + (2m)^2\Pi^2_s - a^2(\Pi_c-\Pi_s)^2]~,\cr
t_3 & = {\ell_A\over R_\alpha (2m)^3 (\Pi_c^2 - \Pi^2_s)}t~,\cr
\phi_3 & = {\alpha\over R_\alpha} + {t_3\over\ell_A}~.
}}
The transformation gives the identifications $\ell_A=2$ and
\eqn\eff{
\rho_\pm = 2mR_\alpha [ m(\Pi_c + \Pi_s)\pm\sqrt{m^2-a^2} (\Pi_c - \Pi_s)]~.
}
These assignments are equivalent to the physical BTZ parameters
\eqn\elc{\eqalign{
M_3 &= {\rho^2_++\rho^2_-\over 8G_3\ell^2_A}= {m^2R^2_\alpha\over 4G_3}
\left( 2m^2 (\Pi^2_c + \Pi^2_s) - a^2 (\Pi_c - \Pi_s)^2\right)~,\cr
J_3 &= {\rho_+\rho_-\over 4G_3\ell_A} = {m^2R^2_\alpha\over 2G_3}
\left( 4m^2 \Pi_c \Pi_s + a^2 (\Pi_c - \Pi_s)^2\right)~.
}}

The effective Newton's constant in three dimensions is determined in terms of the Newton's constant in four dimensions by comparing the reduction from five dimensions on a sphere with radius $\ell_S=1$ to the reduction on a circle with radius $R_\alpha$:
\eqn\ela{
{1\over G_3} = {4\pi\ell_S^2\over G_5} = {4\pi\ell^2_S\over 2\pi R_\alpha G_4}
={2\over R_\alpha G_4}~.
}
This in turn gives the Brown-Henneaux central charge of the effective AdS$_3$
with radius $\ell_A=2$:
\eqn\elb{
c = {3\ell_A\over 2G_3} = {6\over R_\alpha G_4}~.
}
Recall that $R_\alpha$ has dimension of inverse (length)$^2$ so this expression is dimensionless. We are also interested in the effective conformal weights
\eqn\eld{\eqalign{
h_+ &= {M_3\ell_A+ J_3\over 2} = {m^4 R_\alpha\over G_4}(\Pi_c+\Pi_s)^2~,\cr
h_- &= {M_3\ell_A- J_3\over 2} = {m^2 (m^2-a^2) R_\alpha\over G_4}(\Pi_c-\Pi_s)^2~.
}}
Again these expressions are dimensionless because $R_\alpha$ has dimension of (length)$^{-2}$.

The conformal dimensions are generally complicated functions of all physical charges (with implicit dependence on moduli), black hole mass, and black hole angular momentum. However, the combination of charges
\eqn\xc{
I_4= {4m^4\Pi_c \Pi_s\over G_4^2}~,
}
is indepenent of moduli and dependent only on the quantized charges. It is normalized to be an integer. It follows that the effective 3D angular momentum simplifies as
\eqn\xd{
J_3 = h_+ - h_- = {1\over k} ( I_4 + J^2)~.
}
We use the notation $k = c/6$ where $c$ is given in \elb. This is consistent expectations from (generalized) level matching.

The entropy computed from \eld\ by using Cardy's formula gives
\eqn\ele{\eqalign{
S &= 2\pi\left( \sqrt{ch_+\over 6}+\sqrt{ch_-\over 6}\right)\cr
& = {2\pi m\over G_4}\left( (\Pi_c + \Pi_s)m + (\Pi_c -\Pi_s)\sqrt{m^2-a^2}\right)
~.
}}
This agrees with the entropy \ck\ of the original four dimensional black hole, as it
should. 

The agreement for the entropy is not impressive in and by itself. In fact, it 
follows automatically from the local AdS$_3$ structure (for review see \KrausWN). 
In order for a counting to be claimed we must specify the scale $R_\alpha$ which is arbitrary for now. Additionally, we must ascertain that there really is a physical conformal symmetry for which Cardy's
formula \ele\ performs asymptotic state counting. These are the issues we address
in the next section. 

\newsec{Hidden Conformal Symmetry}
There are several promising routes from the facts we have presented to a useful underlying 2D conformal symmetry. In this section we discuss a 4D interpretation (inspired by Kerr/CFT) and a  five dimensional interpretation (generalizing AdS/CFT correspondence). 

\subsec{2D Conformal Symmetry from 4D}
The subtracted geometry with conformal factor \de\ has $SL(2,R)
\times SL(2,R)$ symmetry.
Accordingly, we can represent the scalar Laplacian in the two forms 
\eqn\ya{\eqalign{
\ell^2\nabla^2 &=  {\cal R}_1^2 + {\cal R}_2^2 - {\cal R}_3^2~,\cr
&= {\cal L}_1^2 + {\cal L}_2^2 - {\cal L}_3^2 ~,
}}
where the linear differential operators ${\cal R}_i$ ($i=1,2,3$) and ${\cal L}_i$ ($i=1,2,3$) commute with each other and satisfy $SL(2,R)$ algebras
\eqn\yb{\eqalign{
[{\cal R}_i, {\cal R}_j] = 2i \epsilon_{ijk} (-)^{\delta_{k3}}{\cal R}_k~~;~[{\cal L}_i, {\cal L}_j] = 2i \epsilon_{ijk} (-)^{\delta_{k3}}{\cal L}_k~.
}}

We can construct the differential operators explicitly by comparing with
global AdS$_3$
\eqn\yc{
ds^2_3 = \ell^2 (d\rho^2 - \cosh^2\rho d\tau^2 + \sinh^2\rho d\sigma^2)~.
}
In this standardized setting the Laplacian takes the form \ya\ with the
$SL(2,R)\times SL(2,R)$ generators
\eqn\yd{\eqalign{
{\cal R}_\pm &= {\cal R}_1\pm i {\cal R}_2 = e^{\pm i(\tau+\sigma)} \left(\mp i \partial_\rho +  \tanh\rho\partial_\tau + \coth\rho\partial_\sigma\right)~,\cr
{\cal R}_3 & = \partial_\tau + \partial_\sigma~.
}}
and ${\cal L}_i$ determined by taking $\tau\to -\tau$ in these expressions. 
Comparing the Laplacian in global AdS$_3$
\eqn\ye{
\ell^2\nabla^2 = {1\over \sinh 2\rho}\partial_\rho\sinh 2\rho
\partial_\rho - {1\over\cosh^2\rho}\partial^2_\tau
+ {1\over\sinh^2\rho}\partial^2_\sigma~,
}
and the Laplacian \db\ we find the identifications
\eqn\yf{\eqalign{
\sinh^2\rho &= {r - r_+\over r_+ - r_-}~,\cr
  \sigma - \tau &= - {2\pi i \over\beta_L} ( t - {\beta_R\over \beta_H\Omega_H}\phi)~, \cr
 \sigma + \tau &= - {2\pi i \over\beta_H\Omega_H} \phi~.
}}
These identifications map the $SL(2,R)\times SL(2,R)$ generators in global AdS$_3$ \yd\ to 
the ones adapted to the subtracted black hole background. It is important that the
resulting $SL(2,R)$ generators are canonically normalized 
due to the {\it nonabelian} nature of $SL(2,R)$. In particular the normalized
Cartan generators become 
\eqn\yf{\eqalign{
\pi {\cal R}_3 & = \beta_R i \partial_t + \beta_H \Omega_H i\partial_\phi~,
\cr
\pi {\cal L}_3 & = \beta_L i\partial_t~. 
}}
It may be useful to introduce yet another set of 
coordinates
\eqn\yg{\eqalign{
t^- & = {i\over 2}  (\sigma - \tau) ~, \cr
t^+ & = {i\over 2}  (\sigma + \tau) ~.
}}
such that the generators \yf\ are represented canonically as ${\cal R}_3=i\partial_{t^+}~,~{\cal L}_3=i\partial_{t^-}$. These coordinates generalize the preferred coordinates introduced in \CastroFD\foot{The notation $w^\pm \sim e^{\mp t^\pm}$ was used in \CastroFD. The utility of the $\tau, \sigma$ coordinates was noted already in \CveticUW.}. 

It is a central issue already at the level of the $SL(2,R)\times SL(2,R)$ symmetry that the generators \yd\ are globally ill-defined \CastroFD: the azimuthal angle is identified as $\phi\equiv \phi + 2\pi$ and the generators ${\cal R}_\pm$ and ${\cal L}_\pm$
transform under this identification. It is natural to interpret this ambiguity in terms of a thermal CFT which, because it is a defined on a torus, obeys the equivalences
${\cal R}_\pm \equiv  {\cal R}_\pm e^{-4\pi^2 T_{R}^{\rm CFT}}$, ${\cal L}_\pm \equiv  {\cal L}_\pm e^{-4\pi^2 T_{L}^{\rm CFT}}$. This interpretation determines the relative normalization of the (dimensionful) physical temperatures $T_{R,L}=\beta_{R,L}^{-1}$ and the (dimensionless) CFT temperatures as
\eqn\yh{\eqalign{
T_{L,R}^{\rm CFT} = T_{L,R} \cdot {\beta_R\over\beta_H\Omega_H}~.
}}

To put these values in perspective it is interesting to {\it assume} that the CFT accounts for the black hole entropy \ck\ by satisfying the Cardy formula in the canonical ensemble
\eqn\yi{
S = {\pi^2\over 3} \left( c_L T^{\rm CFT}_L + c_R T^{\rm CFT}_R\right)~.
}
The central charges inferred from this assumption are
\eqn\yj{
c_L = c_R=12\cdot {1\over 4\pi^2} {S_{L,R}\over T_{L,R}}\cdot {\beta_H\Omega_H\over \beta_R}=12J~.
}
It is non-trivial that this procedure gives the same central charge in the $L$ and $R$ sectors (equivalent coincidences were noted in \refs{\KastorGT,\CveticVP}. 

The result \yj\ and the procedure leading to it is a generalization of the hidden conformal symmetry approach \CastroFD\ to the setting with general charges
(see also \refs{\ChenKT\HuangYG\ShaoCF\ChenYU\ChenZWA\ChenAS}). A central weakness of the approach is the {\it assumption} that the $SL(2,R)\times SL(2,R)$ symmetry is enhanced to a Virasoro symmetry (squared) and the {\it assumption} that Cardy's formula applies. It is not presently known how to justify these assumptions. It is nevertheless interesting that the computation suggests a master CFT with the central charge \yi. This value is familiar from the Kerr/CFT correspondence \GuicaMU; but the considerations here suggest (following \CastroFD) that this one theory accounts for the entropy of black holes far from extremality.

Our extension to the setting with arbitrary charges creates a tension between this optimistic interpretation of the Kerr/CFT correspondence and the standard description (such as \MaldacenaDE) that applies near the BPS limit. Such ``large charge'' descriptions invoke CFT's with a central charge that depends on 
spacetime charges and in many cases these CFT's describes black holes with a 
range of the angular momenta. It would be interesting to delineate the range of 
applicability of these disparate descriptions. 

\subsec{2D Conformal Symmetry from 5D}
Our embedding of the subtracted geometry into five dimensions suggests a different approach to the apparent conformal symmetry: the local AdS$_3\times S^2$ invites reference to standard AdS/CFT correspondence \MaldacenaRE\ or, 
more precisely, the Brown-Henneaux result \BrownNW. Concretely, we can map the BTZ black hole \efd\ into global AdS$_3$ \yc\ through the identifications 
\eqn\xa{\eqalign{
t^\pm & = {i\over 2}  (\sigma \pm \tau) = {r_+ \pm r_-\over 2\ell_A} z^\pm
~, 
}}
where, according to the embedding \efe,
\eqn\xb{\eqalign{
z^-&=\phi_3 - {t_3\over\ell_A}= {\alpha\over R_\alpha}~,\cr
 z^+ &= \phi_3 + {t_3\over\ell_A}={\alpha\over R_\alpha} + {2\ell\over R_\alpha (2m)^2 (\Pi_c^2 - \Pi^2_s)}t~.
 }}
An obvious advantage of this approach is that the $SL(2,R)$'s do in fact extend to full Virasoro's: diffeomorphisms that act on $z^\pm$ (while preserving asymptotic AdS$_3$) form a Virasoro algebra in AdS$_3$. A key issue then becomes the value of the central charge, given previously in \elb. It is 
determined entirely by $R_\alpha$, the periodicity of the auxiliary dimension. This parameter can be inferred from the fibration of the sphere $S^2$ over BTZ: according to \ef\ it is only the combination $\alpha+2G_4J\phi$ that enters so the azimuthal shift symmetry $\phi\to\phi+2\pi$ is equivalent to 
$R_\alpha=2G_4J$. The central charge \elb\ then returns to the Kerr value $c=12J$. 

Although the 5D interpretation thus appears to have the same central charge as the 4D interpretation, the states in the Virasoro representation are quite different: in 5D the states generally depend on the coordinate $\alpha$. An added value of the 5D representation is that it realizes modular invariance (and spectral flow symmetry) in a simple manner. This ensures that conformal symmetry acts on a sufficient number of conformal primaries that the black hole entropy is accounted for, rather than just on the AdS$_3$ vacuum. It would be interesting to understand the relations between the 4D and 5D interpretations. 

There is another interesting periodicity that is determined by the set-up. Recall that the azimuthal angle $\phi$ plays the role of time near the horizon and regularity of the Euclidean geometry fixes its imaginary periodicity as \ce. This periodicity is computed with a combination of $\phi$ and the asymptotic time $t$ fixed, and this in turn determines the imaginary periodicity of $t$ as \cf. These shifts are with $\alpha$ fixed but \ef\ show that they are equivalent to the imaginary periodicity 
\eqn\xe{
\beta_\alpha = {\beta_\phi\over 2ma (\Pi_c - \Pi_s)} = {2\pi\over 2m(\Pi_c  -\Pi_s)\sqrt{m^2-a^2}}~.
}
This quantity can be interpreted as usual as the chemical potential for excitations with momentum along the auxiliary direction $\alpha$. It can be expressed geometrically as
\eqn\xf{
{1\over\beta_\alpha} = {A_+-A_-\over 16\pi^2}~,
}
where $A_\pm$ are the areas of the outer and inner horizon. The geometric nature of this formula suggests a robust significance of this potential. We defer further exploration of its origin to future work.
%

\bigskip
\noindent {\bf Acknowledgments:} \medskip \noindent
We thank G. Comp\`{e}re,  M. Guica, C. Keeler, P. Kraus,  C. Pope,  and especially G. Gibbons
for discussions. We thank the Aspen Center for Physics for hospitality. MC is supported by the DoE 
Grant DOE-EY-76-02- 3071, the NSF RTG DMS Grant 0636606, the Fay R. 
and Eugene L. Langberg Endowed Chair  and the Slovenian
Research Agency (ARRS).
FL is also supported by the DoE.

\listrefs
\end